\documentclass{appolb}
\usepackage{graphicx}
\usepackage{amsmath}
\usepackage{cite}
\usepackage{color}

\begin{document}
\title{Modelling of the European Union income distribution by extended Yakovenko formula 
}
\author{Maciej Jagielski\footnote{e-mail: zagielski@interia.pl}, Ryszard Kutner
\address{Institute of Experimental Physics,\\ Faculty of Physics, University of Warsaw,\\ Ho\.za 69, PL-00681 Warszawa, Poland}
\\
}
\maketitle
\begin{abstract}
We found a unified formula for description of the household incomes of all society classes, for instance, for the European Union in years 2005-2010. The formula is more general than well known that of Yakovenko et al. because, it satisfactorily describes not only the household incomes of low- and medium-income society classes but also the household incomes of the high-income society class. As a stricking result, we found that the high-income society class almost disappeared in year 2009, in opposite to situation in remaining years, where this class played a significant role.
\end{abstract}
\vspace{0.5cm}
SUBMITTED TO ACTA PHYSICA POLONICA B  
\section{Introduction}
One of the major trends in econophysics is the study of the redistribution of wealth and income in society, and the analysis of social inequalities.  A pioneer of this type of research is the Italian economist and sociologist -- Vilfredo Pareto \cite{RHCR_2006, M_1960, M_1963}. He found that the distribution functions of individual incomes in different countries (within stable economy) could not resemble the distribution functions obtained if gain and accumulation of income were random. Pareto also analysed the stability of these distributions. That is, he found that even if one removes from the society structure the richest or poorest members of the society, after a certain period of time, the income distribution function will be again completed in the form almost the same as the initial distribution function \cite{P_1897, M_1960, RHCR_2006}. The main result of Pareto's analysis was the conclusion that distribution functions of income in each country (with a stable economy) can be described by a universal power-law, known nowadays as the Pareto law. As a possible origin of this law, Pareto indicated a self-similarity structure of societies.

The income of societies was also analysed by Robert Gibrat \cite{G_1931, K_1945, A_1995, S_1997, RHCR_2006}, David Champernowne \cite{Ch_1953}, and Benoit Mandelbrot \cite{M_1960, M_1963}.
Their studies led to the disclosure of many important properties of income distributions, however, did not give a satisfactory answer to crucial question concerning the microscopic (microeconomic) mechanism determining the empirical complementary distribution functions.

Several models, trying to explain the microscopic mechanisms (for income dynamics of individuals or households) standing behind the empirical complementary distribution functions of income, were proposed. These models consider the income of individual or household as a random variable. To describe the dynamics of this variable, the nonlinear stochastic Langevin equation and the corresponding Fokker--Planck equation are used as a natural foundation. Upon the specific assumptions concerning the dynamics of income, we can obtain the following models: (i) the Boltzmann--Gibbs law \cite{RHCR_2006, YR_2009, BY_2010}, (ii) the Pareto law \cite{RHCR_2006, YR_2009, BY_2010}, (iii) the Rule of Proportionate Growth \cite{G_1931, K_1945, A_1995, S_1997, RHCR_2006, YR_2009}, (iv) the Generalised Lotka--Volterra model \cite{SR_2001, RS_2001, SR_2002, H_2004, RHCR_2006, YR_2009}, and (v) Yakovenko et al. model \cite{YR_2009, BY_2010}.

However, none of the models developed so far (best of our knowledge) gives an analytical description of the annual household incomes of all society classes (i.e. the low-, medium-, and high-income society classes) by a single unified formula based on a unified formalism. In the present paper we extend and complement the results of our recent model \cite{JK_2013a, JK_2013b} with the low number of free parameters that 
well reproduces the empirical complementary cumulative distribution functions. As the most striking result which we found is almost total decay of the high-income society class in 2009, while in all other years in this century the high-income society class is quite robust against the financial markets turbulence.  

\section{Data description}
We used empirical data from Eurostat's Survey on Income and Living Conditions (EU--SILC) \cite{EURO, EURO_2005, EURO_2006, EURO_2007, EURO_2008, EURO_2009, EURO_2010} for years 2005--2010. This database contains general information on the demographic characteristics of households in the European Union (EU), their living conditions, income and economic activity. We chose to our analysis the variable \emph{Total household gross income}. Eurostat's EU--SILC data contain only few observations on the households belonging to high-income society class. This means that they cannot be subjected to any statistical description. Therefore, in order to consider the high-income society class, we additionally analysed the effective income of billionaires in the EU by using the Forbes "The World's Billionaires" rank\footnote{The term "billionaire" used herein is equivalent (as in the US terminology) to the term "multimillionaire" used in the European terminology. Since we consider wealth and income of billionaires in euros, we recalculated US dollars to euros by using the mean exchange rate at the day of construction of the Forbes "The World's Billionaires".}$^{,}$\footnote{The billionaires who gained effective incomes are billionaires whose incomes are greater than zero.} \cite{FORBES}.

Using EU--SILC database and rank of the richest Europeans we were able to consider incomes of three society classes thanks to the procedure presented in Ref. \cite{JK_2013b}. Hence, we received the data record sufficiently large for statistical study of all society classes, including the high-income society class. Notably, in our studies we analysed the empirical complementary cumulative distribution function by using the well known Weibull rank formula \cite{CMM_1988, Han_2004}. 


\section{Extended Yakovenko et al. model}

Let $m$ be an influx of income per unit time to a given household. We treat $m$ as a variable obeying stochastic dynamics. Then, we can describe time evolution of probability distribution function of income by using the Fokker-Planck equation
\begin{eqnarray}
\frac{\partial}{\partial t}P(m,t)&=&\frac{\partial}{\partial m}[A(m)P(m,t)]+
\frac{{\partial ^2}}{\partial m^2}\left[B(m)P(m,t)\right]. 
\label{rown2a}
\end{eqnarray}
where, $B(m)=C^2(m)/2$ and $P(m,t)$ is the temporal income distribution function. In general, functions 
$A(m)$ and $B(m)$ can be additionally determined by the first and second moments of the income change per unit time, respectively, only if these moments exist. Subsequently, the equilibrium solution of Eq. (\ref{rown2a}), $P_{\rm eq}$, takes the form \cite{vanK_1990}:
\begin{eqnarray}
P_{\rm eq}(m)=\frac{const}{B(m)}\exp\left(-\int_{{m_{\rm init}}}^m\frac{A(m')}{B(m')}\, dm'\right),\; 
\frac{const}{B(m)}>0,
\label{rown5}
\end{eqnarray}
where integral should be a non-negative quantity, $m_{\rm init}$ is the lowest household income, and $const$ is a normalisation factor. Fortunately, both It\^o and Stratonovitch representations \cite{vanK_1990} give almost the same equilibrium distribution function. Eq. (\ref{rown2a}) and its equilibrium solution, Eq. (\ref{rown5}), define the formalism of the income change, which remains the same for the whole society.

By using Eq. (\ref{rown5}) we are able to derive such a distribution function which would cover all three ranges of the empirical data records, i.e. low-, medium-, and high-income classes of the society (including also two short intermediate regions between them). To make it, we have to provide function $A(m)$ in a threshold form \cite{JK_2013a, JK_2013b}: 
\begin{eqnarray}
A(m) &=& \left\{ \begin{array}{ll}
A^<(m)=A_0+a\, m & \textrm{if $m<m_1$} \\
{A^{\ge }(m)=A'_0}+a'\, m & \textrm{if $m\ge m_1$,}
\end{array} \right. \nonumber\\
B(m)&=&B_0+b\, m^2=b\, (m^2_0+m^2),
\label{rown12}
\end{eqnarray}
where parameters used in these relations are defined and considered below.

The form of $A(m)$ and $B(m)$ given by (\ref{rown12}) allows the coexistence of additive and multiplicative stochastic processes. Thus, we assume that household income consists of two components. First -- the deterministic component of income arises from the fact that household income can take the form of wages and salaries. Second -- indeterministic component may express profits which go to household mainly through investments and capital gains.

At the threshold $m_1$, there is a jump only of the proportionality coefficient of the drift term that is, this coefficient abruptly changes from $a$ to $a'$ (as $a\ne a'$), while $P_{\rm eq}(m)$ has no discontinuity there.

The threshold parameter $m_1$ is interpreted as a crossover income between the medium- and high-income society classes and parameter $m_0$ is the crossover income between the low- and medium-income society classes. 

By substituting Eq. (\ref{rown12}) into Eq. (\ref{rown5}), we get two-branch distribution function \cite{JK_2013a, JK_2013b}
\begin{eqnarray}
P_{\rm eq}(m) =\left\{ \begin{array}{ll}
c'\, \frac{\exp\left(-(m_0/T)\arctan(m/m_0)\right)}{[1+(m/m_0)^2]^{(\alpha +1)/2}}, & \textrm{if $m<m_1$} \\
c''\, \frac{\exp\left(-(m_0/T_1)\arctan(m/m_0)\right)}{[1+(m/m_0)^2]^{(\alpha_1 +1)/2}}, & \textrm{if $m\ge m_1$}
\end{array} \right. 
\label{rown19}
\end{eqnarray}
where exponents $\alpha=1+a/b$, $\alpha _1=1+a'/b$, and income temperatures $T=B_0/A_0$, $T_1=B_0/A'_0$. Parameter $T$ can be interpreted in this case as an average income per household for low-income society class. Parameter $T_1$ has the same interpretation but for high-income society class. Apparently, the number of free (effective) parameters driving the two-branch distribution function, given by Eq. (\ref{rown19}), is reduced because this function depends only on the ratio of some parameters defining the Fokker-Planck equation given by Eq. (\ref{rown2a}).

\section{Results}
We compared the theoretical complementary cumulative distribution function based on our probability distribution function $P_{\rm eq}(m)$, given by Eq. (\ref{rown19}), with the empirical data for the whole income range. However, the analytical form of this theoretical complementary cumulative distribution function is unknown in the closed explicit form. Therefore, we calculated it numerically for each value of income $m$. 

\begin{figure}[htb]
\centerline{
\includegraphics[scale=0.45]{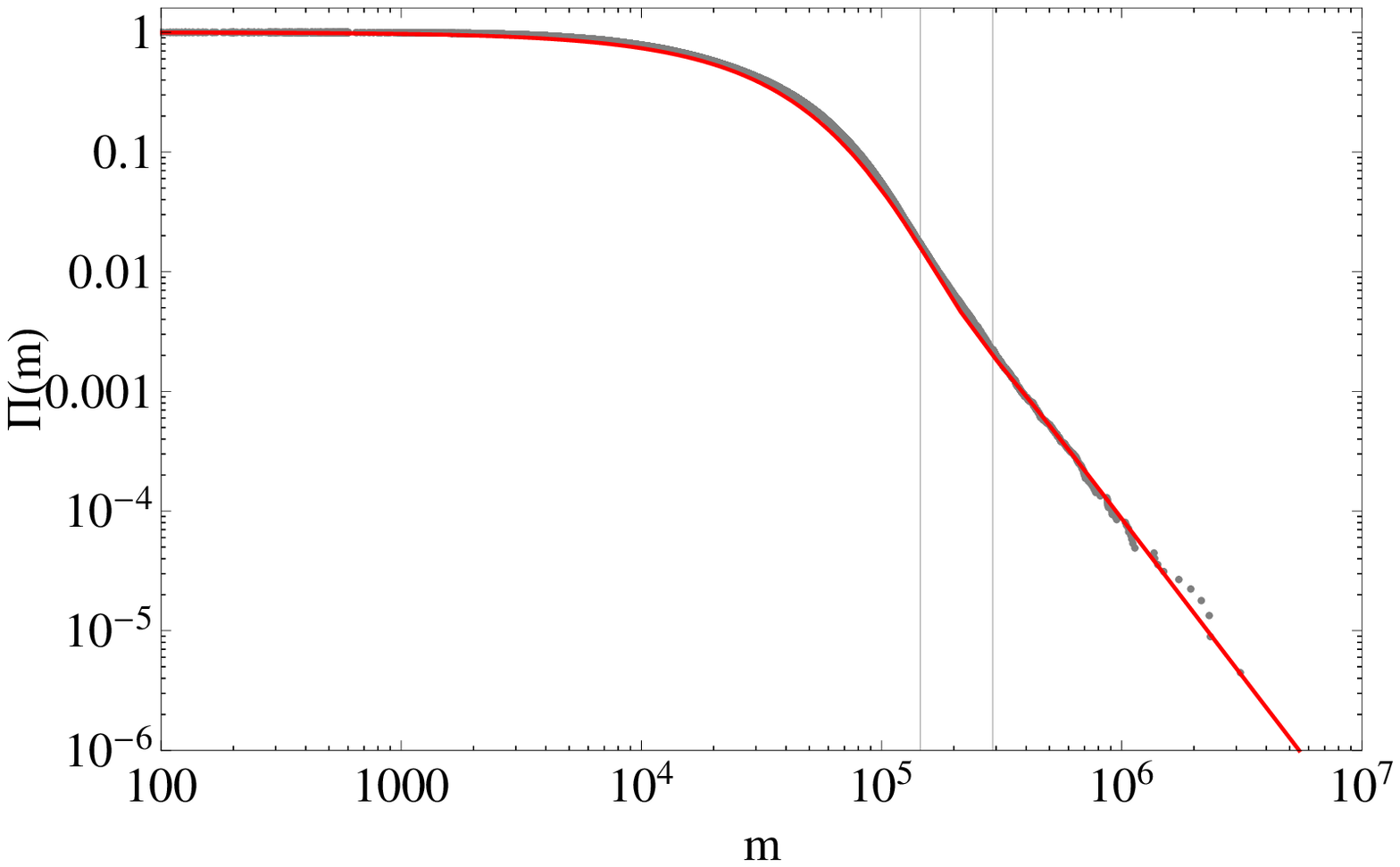}}
\caption{Comparison of the complementary cumulative distribution function, based on the Extended Yakovenko 
et al. formula, Eq. (\ref{rown19}), (solid line) with the EU household income empirical data set (dots) for year 2009 ($T=37\times 10^3\pm 3\times 10^3$ EUR, $T_1=2.9\times 10^5\pm 0.5\times 10^5$ EUR$, m_0=1.45\times 10^5\pm 0.20\times 10^5$ EUR, 
$m_1=2.9\times 10^5\pm 0.5\times 10^5$ EUR, $\alpha = 2.974 \pm 0.001$, and $\alpha_1 = 2.608 \pm 0.006$). The first and the second vertical lines are placed at $m_0$ and $m_1$, respectively \cite{EURO_2009, FORBES}. Apparently, herein only low- and medium-income society classes are present, while high-income society class is, in practise, absent in this year.}
\label{fig1}
\end{figure}

\begin{figure}[htb]
\centerline{
\includegraphics[scale=0.45]{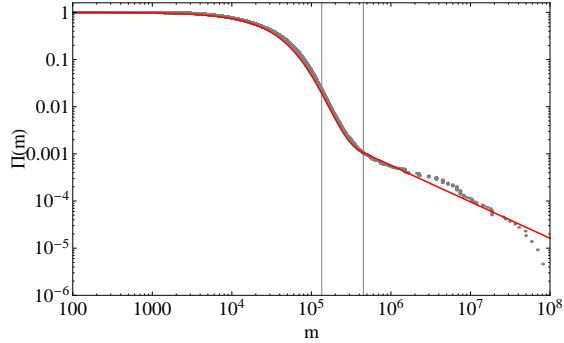}}
\caption{Comparison of the complementary cumulative distribution function, based on the Extended Yakovenko et al. formula, Eq. (\ref{rown19}), (solid line) with the EU household income empirical data set (dots) for year 2010 ($T=38\times 10^3\pm 3\times 10^3$ EUR, $T_1=4.5\times 10^5\pm 0.5\times 10^5$ EUR, $m_0=1.35\times 10^5\pm 0.20\times 10^5$ EUR, 
$m_1=4.5\times 10^5\pm 0.5\times 10^5$ EUR, $\alpha = 3.153 \pm 0.002$, and $\alpha_1 = 0.77 \pm 0.01$). The first and the second vertical lines are placed at $m_0$ and $m_1$, respectively \cite{EURO_2010, FORBES}. Apparently, all income society classes are present in this year.}
\label{fig2}
\end{figure}

The corresponding plots of the empirical and theoretical complementary cumulative distribution functions in the log-log scale are plotted in Fig. \ref{fig1} and Fig. \ref{fig2}, for instance, for years 2009 and 2010, respectively. In addition, the Table \ref{tab1} provides estimates of the parameters of the Extended Yakovenko et al. model for years 2005--2010; the errors of parameters are given in Table \ref{tab2}.

Apparently, the predictions of the Extended Yakovenko et al. formula, Eq. (\ref{rown19}), (solid curve in Fig. \ref{fig1} and Fig. \ref{fig2})  well agree with the empirical cumulative distribution functions of annual total gross income of households in the European Union (dots in Figs. \ref{fig1} and \ref{fig2}). Thus, using this formula, we can describe the income of all three society classes, namely the low-, medium- and high-income society classes. Such a good agreement with the empirical data was obtained primarily through the adoption of the following significant assumptions:
\begin{itemize}
\item Extended Yakovenko et al. model allows for the coexistence of additive and multiplicative processes and differentiates detailed dynamics of income. It is assumed, in this model, that low- and medium-income society classes gain or lose income differently than the high-income society class,
\item Extended Yakovenko et al. model satisfies the continuity condition of the probability distribution function of income. However, it does not exclude the possibility that some values of parameters can be discontinuous.
\end{itemize}

\begin{table}[htb]
\centering
\caption{Parameters obtained from the comparison of the Extended Yakovenko et al. model with empirical cumulative distribution functions of the annual total gross income of households in the European Union for years 2005--2010. \vspace {0.5cm}}
 \begin{tabular}{|c|c|c|c|c|c|c|}
  \hline 
 {\bf Year} & $\boldsymbol{T}$ & $\boldsymbol{m_0}$ {\bf [EUR]} & $\boldsymbol{\alpha}$ 
& $\boldsymbol{T_1}$ & $\boldsymbol{m_1}$ {\bf [EUR]} & $\boldsymbol{\alpha_1}$\\
\hline
	2005 & $36\,000$ & $155\,000$ & 2.907 & 
$430\,000$ & $430\,000$ & 0.795\\
  \hline
	2006 & $37\,000$ & $145\,000$ & 2.892 & $445\,000$ & $445\,000$ & 0.86\\
  \hline
2007 & $37\,000$ & $160\,000$ & 2.735 & $480\,000$ & $480\,000$ & 0.79\\
  \hline
	2008 & $38\,000$ & $120\,000$ & 2.965 & 
$450\,000$ & $450\,000$ & 0.890\\
  \hline
	2009 & $37\,000$ & $145\,000$ & 2.974 & $290\,000$ & $290\,000$ & 2.608\\
  \hline
2010 & $38\,000$ & $135\,000$ & 3.153 & $450\,000$ & $450\,000$ & 0.77\\
  \hline
\end{tabular} 
\label{tab1}
\end{table}

\begin{table}[htb]
\centering
\caption{The errors of the model parameters obtained from the comparison of the Extended Yakovenko et al. model with empirical cumulative distribution functions of the annual total gross income of households in the European Union for years 2005--2010. \vspace {0.5cm}}
 \begin{tabular}{|c|c|c|c|c|c|c|}
  \hline 
 {\bf Year} & $\boldsymbol{\Delta T}$ & $\boldsymbol{\Delta m_0}$ {\bf [EUR]} & $\boldsymbol{\Delta \alpha}$ 
& $\boldsymbol{\Delta T_1}$ & $\boldsymbol{\Delta m_1}$ {\bf [EUR]} & $\boldsymbol{\Delta \alpha_1}$\\
\hline
	2005 & $3\,000$ & $20\,000$ & 0.003 & 
$50\,000$ & $50\,000$ & 0.009\\
  \hline
	2006 & $3\,000$ & $20\,000$ & 0.004 & $50\,000$ & $50\,000$ & 0.01\\
  \hline
2007 & $3\,000$ & $20\,000$ & 0.004 & $50\,000$ & $50\,000$ & 0.01\\
  \hline
	2008 & $3\,000$ & $20\,000$ & 0.001 & 
$50\,000$ & $50\,000$ & 0.007\\
  \hline
	2009 & $3\,000$ & $20\,000$ & 0.001 & $50\,000$ & $50\,000$ & 0.006\\
  \hline
2010 & $3\,000$ & $20\,000$ & 0.002 & $50\,000$ & $50\,000$ & 0.01\\
  \hline
\end{tabular} 
\label{tab2}
\end{table}

It is seen from Table \ref {tab1} that income temperature $T$ is only a slowly varying function of time. 

Apparently, parameter $m_0$ (cf. Table \ref {tab1}) oscillates around a mean value $143\,333$ Euro. This parameter can be considered as a crossover income between low- and medium-income society classes. Similarly the parameter $m_1$, which for the years 2005--2008 and 2010 oscillates around mean value $451\,000$ Euro (except for year 2009), is a crossover income between medium- and high-income society classes.

Changes in the value of exponent $\alpha$ (cf. Table \ref{tab1}) show that for the time period 2005--2007, this exponent has been declining and beginning from year 2007 monotonically increases. This means that in years 2005--2007 social stratification within the medium-income society class has increased, and then in years 2008--2010 -- decreased. However, the parameter $\alpha_1$ in years 2005--2008 and in year 2010, changed slightly. In other words, social stratification within the high-income society class remained, more or less, at the same, high level (i.e. having $\alpha_1 < 1$).

For year 2009, we observed that values of parameters $T_1$, $m_1$, and $\alpha_1 $ differ significantly in comparison with remaining years. In this year, there was a rapid decrease in incomes of the high-income society class, as well as the huge decrease in the number of households belonging to this society class. This situation was due to the economic crisis which began in year 2007, and its peak was in year 2008 (affecting the income received in the following year, i.e. 2009). The crisis resulted in a much lower value of crossover income, $m_1$. This also contributed to a significant reduction in the social stratification within the high-income society class (in year 2009 there was a significant increase of parameter $\alpha_1$ up to value $2.608$), actually, making this class the member of medium-income society class. Therefore, it can be noted that in year 2009 the high-income society class, in principle, does not exist.   
However, for the years beyond the year 2009, the shape of empirical complementary cumulative distribution functions is quite persistent. We can only notice a change in the number of households belonging to specific society classes but the income structure of society (as a whole) remains basically unchanged.

\section{Concluding remarks}
We suppose that some parameters of Extended Yakovenko et al. formula, Eq. (\ref{rown19}), play a role of indicators of crisis. For instance, the crisis does not affect low-income society class (parameter $T$ practically does not change), but leads to the lower social stratification within the medium-income society class. It should be noted that in the case of the high-income society class exponent $\alpha_1$ experiences a rapid increase. Thus, the financial impact of the crisis on the high-income society class in the European Union was extremely severe. It seems, that only the analysis of incomes of medium- and high-income society classes may give an answer to the crucial question whether the crisis is coming. However, in order to reach some definite, deeper conclusions (especially on the universality of the obtained results) further study is required, involving systematic comparisons with previous crises. 

We believe that our results will contribute to a better understanding of the mechanisms of enrichment and impoverishment of households, social classes and whole societies as well. It is also very likely that we find quite precise classification of income ranges which determine whether the household belongs to the low-, medium- or high-income society class. Values of parameters obtained from comparison of our theoretical model, Eq. (\ref{rown19}), with empirical data can be used to define advanced indicators of social inequalities. In economics to measure social inequality the Gini coefficient is used \cite{G_1936, DWMW_1987, DWMW_1988, DW_2000}. We proposed an alternative approach, which besides more sensitive indicators of social inequalities, offers valuable theoretical explanation on a level of microscopic dynamics of individual household's income.


\bibliographystyle{unsrtmy}
\bibliography{bibliography}
\end{document}